# Noise-resistant reconstruction algorithm based on the sinogram pattern

Byung Chun Kim[1], Hyunju Lee[1], Kyungtaek Jun[1,*]

[1] Research center, Innovation on Quantum and Computed Tomography, Seoul, 04364, Republic of Korea
* Correspondence: ktfriends@gmail.com

## Introduction

    X-ray computed tomography (CT) is an imaging technique that can obtain the structure of an object through the penetration of X-rays at different projection angles. CT is an essential technique used in many fields such as biology, archaeology, geoscience, and materials science because it can obtain the structure of a sample without destroying the object [1-6]. In addition, many CT examinations are performed annually for the purpose of medical diagnosis [7,8]. Despite the importance of CT, there are still several difficulties in obtaining a clean CT image due to limitations of the back-projection algorithm [7], dissatisfaction with the Beer-Lambert law [8], and various artifacts [9-14].

    Types of CT scan can be broadly divided into spiral CT [15], electron beam tomography [16], and synchrotron x-ray tomography [17]. Each CT system has been developed to suit the size and characteristics of the sample. In synchrotron x-ray tomography, a backprojection method based on fast Fourier transform (FFT) has been developed [18]. A filter algorithm was also developed to obtain a CT image like the actual structure of the sample [19]. FFT, which is now used as a backprojection algorithm in synchrotron x-ray tomography, is faster than iteration-based algorithms and reconstruct clean CT images [20]. In particular, even if the sample size is larger than the charged couple device, it can reconstruct a clean CT image. However, it has the disadvantage of being vulnerable to ring artifacts caused by CCD errors. So, various algorithms for ring artifact correction have been developed [10,21,22,23]. However, ring artifacts are still a problem in reconstructing clean CT images.

    In this paper, we introduce a new CT image reconstruction algorithm that is less affected by various artifacts. The new reconstruction algorithm is a method of minimizing the difference between synchrotron X-ray tomography data and sinograms generated using Radon transform of CT images. The CT image is iteratively updated to reduce the difference from the sinogram of the data. This method can obtain clean CT images from the projection data, which can create ring artifacts or metal artifacts. Also, even if the sample size is larger than the CCD and/or the projection image does not satisfy the Beer-Lambert law, a clean CT image can be reconstructed. Our new reconstruction algorithm can also be applied to fan beam CT or cone beam CT.

## Method

**Reconstruction algorithm based on sinogram**

To discuss our reconstruction algorithm, we first look at the sinogram related to axial level of the sample. A sinogram consists of the sum of values related to the x-ray mass attenuation of all points in the sample. In general, the concentration-absorbance graph does not satisfy the Beer-Lambert law and has a negative deviation graph. Especially when the sample contains metal, the CT image will have lower pixel values inside the sample [11]. However, if the size of the metal is smaller than the total size of the sample, the density has a sufficiently good value for the area where the x-ray does not penetrate the metal. Our algorithm assumes that the sinogram of a sample is generally reliable even in the presence of various artifacts. The algorithm iteratively approaches the CT image that makes the most similar sinogram to the sinogram of the sample obtained through X-ray. When it is necessary to determine the center of rotation in the sinogram, it is padded on one side of the sinogram to preserve as much sample information as possible. New CT image reconstruction algorithm can be formulated as an optimization problem as follow:

$$\underset{T}{\operatorname{argmin}} \operatorname{MSE}(R(T), S)$$

where $S$ is the given sinogram and $R$ is the Radon transformation.

If we set S as X-ray projection images and extend T as 3D image, we can easily get a reconstruction algorithm for CBCT. But in this paper, we focus on the 2D image cases, because of the limitation of computing resources.

We wrote a Python (3.9.5) code which solves the optimization problem using the gradient descent method. To hit this goal, we had to write a differentiable Radon transformation using PyTorch (1.9.0+cu111), and we used the Adam optimizer built in PyTorch library. We utilize NVIDIA Tesla V100 SXM2 GPU for acceleration.

Reconstruction process separated in two steps:

First step: we draw a draft of reconstructed CT image. We update T 2000 times using Adam optimizer with learning rate 0.01, and blur T using the Gaussian filter with $\sigma = 5$.

Second step: we update T 2000 more times.

**Reconstruction algorithm including ring artifact removal algorithm**

Suppose a few rows in the sinogram completely lose information about the density of the sample. In the dead region of the sinogram, information corresponding to the density is lost, but each region of the sample has a sinusoidal trajectory, so it has information when it leaves the dead region. In this case, our reconstruction algorithm can be formulated as an optimization problem without the dead region.

Let S be a (m× n) image matrix and let M be a (m× n) Boolean matrix which is called a mask of S. SM denotes the part of S which marked as True in M, called the masked image. The mean squared error of two masked images using the same mask is defined by the mean squared error of the elements in the images marked as True in M. New CT image reconstruction algorithm can be formulated as a marked optimization problem for the form as follow:

$$\underset{T}{\operatorname{argmin}} \operatorname{MSE}(R(T)_M, S_M)$$

where $S$ is the given sinogram and $M$ be the mask which marks the missing data part as False. In the case of the sinogram contains noise, we can deploy the dropout by putting some random Falses in the mask $M$.

**Removing the outside of the ROI**

We introduce a way of cleaning a sinogram. In a CT image T, each point represents the X-ray mass attenuation coefficient, and the corresponding sinogram can be thought as the sum of sine curves corresponding to the points of T. Let N be a mask of T which represents the outside of the ROI. Then $R(T) - R(T_N)$ gives the cleaned up sinogram which corresponding to the ROI of T.

For real world problem case, the first part of the reconstruction process is the same as before except to use of cleaned up and masked sinogram. But in the second stage, we update T 5000 more times with small learning rate, 0.001, and use 90% dropout to prevent overfitting.

**Result**

An image sample and ideal sinograms were used to measure the results and errors of the new reconstruction algorithm. The image sample used is a Shepp-Logan phantom (Fig. 1a). The sample size is 400*400 and 360 projection images for 180 degrees were generated using a Radon transform (Fig. 1b). The region of interest (RoI) for calculating the mean squared error (MSE) is shown in Fig. 1(c).

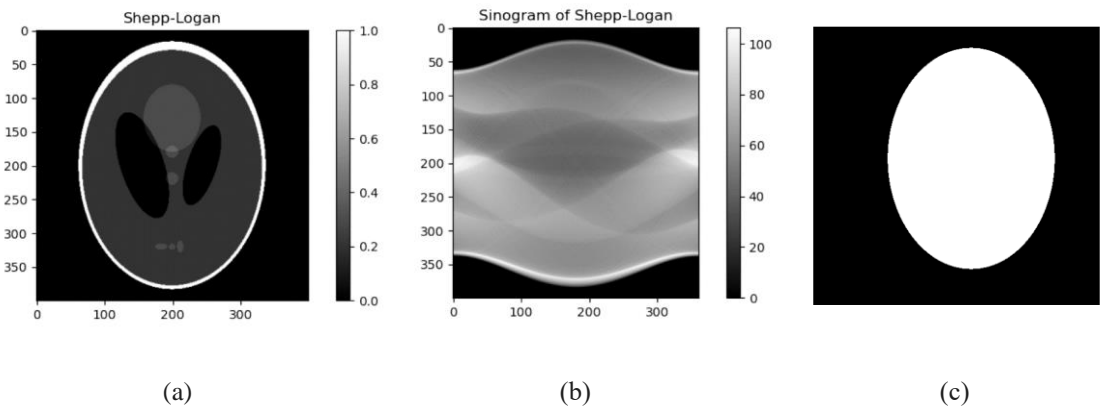

(a)　　　　　　　　　　　(b)　　　　　　　　　　　(c)

Figure 1. Test image sample, its sinogram, and RoI

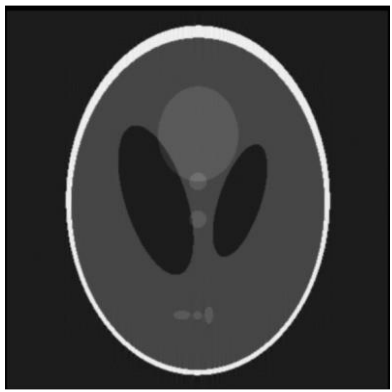

(a)

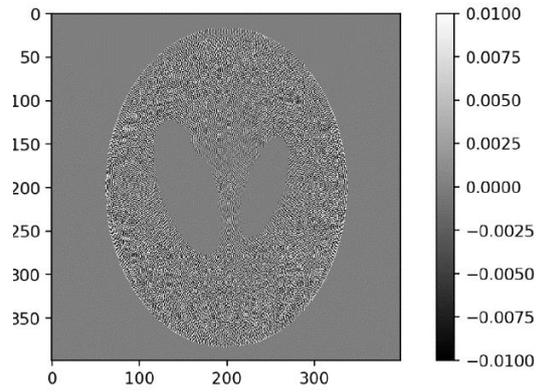

(b)

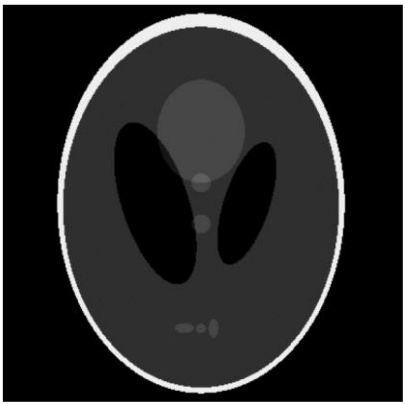

(c)

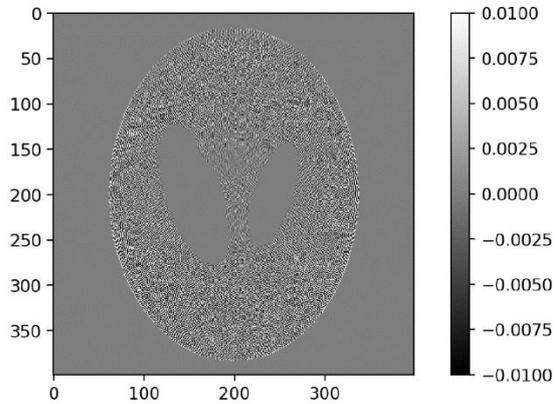

(d)

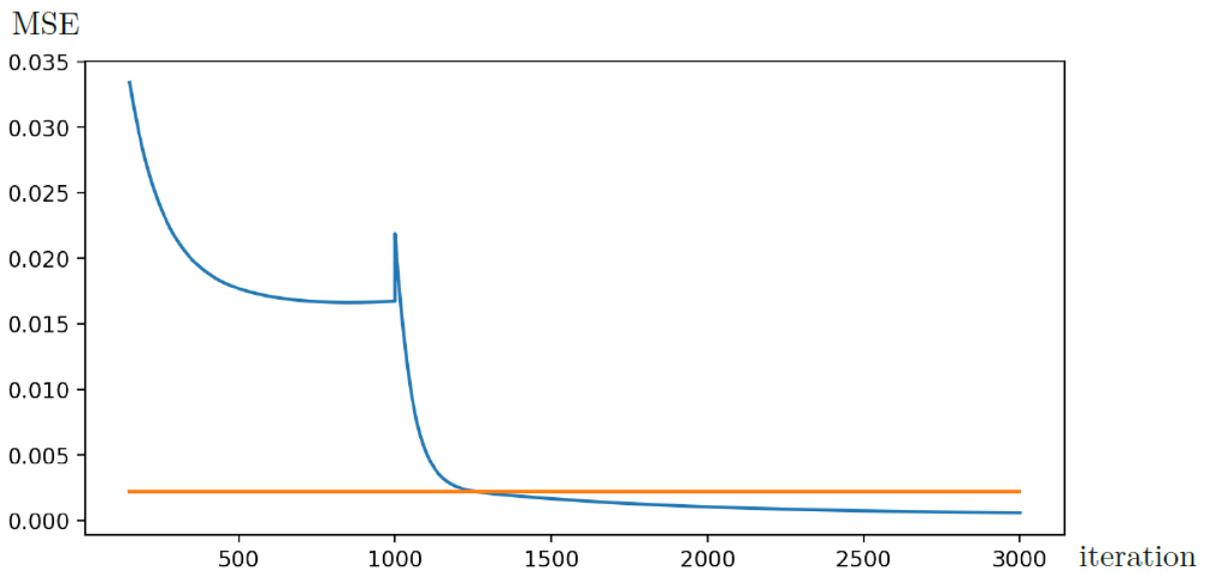

(e)

Figure 2. Result comparison. (a) Reconstructed image from a sinogram in Fig. 1b using the FFT. (b) Difference

between Figs. 1a and 2a inside the ROI. (c) Reconstructed image from a sinogram in Fig. 1b using our algorithm. (d) Difference between Figs. 1a and 2c inside the ROI. (e) A MSE plot about colored curves: the FFT algorithm (--) and out algorithm (--).

Figure 2a is the image produced by the Fast Fourier Transform (FFT), and Fig. 2b shows the difference within the RoI for Figs. 1a and 2a. Figures 2c and d show the images and differences produced by our algorithm, respectively. The reconstructed image created using our optimization method increases the convergence speed by applying a Gaussian filter at 1000 iterations. If the iteration exceeds 1250 times, it can be seen that the MSE of our method is smaller than that of the existing FFT method. The MSE of the FFT is about 0.0022, and the MSE of our method is about 0.00048. Our method reconstructs the image closer to the real image than the conventional method.

Figure 3a occurs when the density value cannot be accurately obtained due to a malfunction of the CCD during scanning. We are using a test sinogram sample that completely loses density information around the 250th pixel. In such cases, ring artifacts will appear in the reconstruction image in Fig. 3b. In this case, we apply the optimization method except for the dead rows of the sinogram in Fig. 3a. The results obtained with the ring artifact resistant reconstruction algorithm are shown in Fig. 3c. The MSE for the reconstructed image in Fig. 3c is 0.00050. This value is lower than the MSE of the reconstructed image obtained from an error-free sinogram using FFT. The image in Fig. 3d shows the difference between Figs. 1a and 3c.

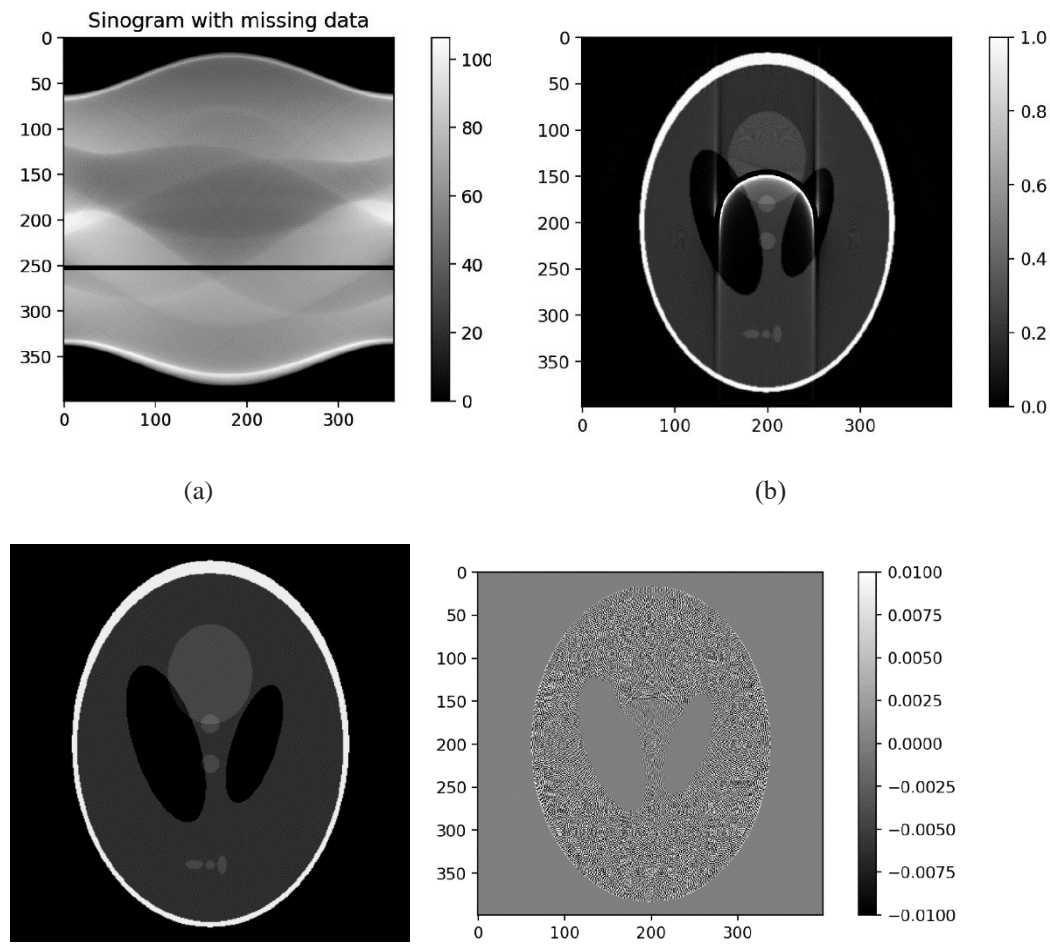

(c) (d)

Figure 3. Ring artifact resistant reconstruction algorithm based on the sinogram pattern

## Discussion

Until now, studies to correct completely dead pixels in sinograms have mainly used interpolation using the perimeter of dead pixels. Although this method lacks physical evidence, it has made it possible to produce better CT images. Our new method uses a 2D or 3D sinogram to find the x-ray mass attenuation coefficient for each grid of the sample. The new algorithm can produce good results even if there are various artifacts if you can be trusted to some extent in the projection image set. To prove this, we used a set of projection images of samples containing metals. Because metal only appears in projection images for specific projection angles, projected information of the metal about the full 180-degree angle is lacking. In addition, sinograms with dead pixels that create ring artifacts in high-quality CT images were used. Our new algorithm uses the property that the trajectory of the x-ray mass attenuation coefficient of each grid of the sample moves along the trajectory of the sinusoidal function in the sinogram. So, even if there is no information about the density of the sample in a certain part of the sinogram or it is incorrect, it is possible to find as much similar information as possible in the other part. It minimizes artifacts as much as possible and reconstructs clean CT images.

Another advantage of our reconstruction algorithm is that it can be applied to fan beam CT or cone beam CT systems. The cone beam and fan beam projection geometries are already well known. Our algorithm is applicable to all CT systems if the projection geometry can be defined. The projection used in the new algorithm satisfies the Beer-Lambert law because it uses a Radon transform. Since these results are different from the actual density, some errors may be created in the CT image. If we can incorporate all the physical properties that x-rays can produce as they penetrate the sample into our new reconstruction algorithm, we can get better CT images. Also, this algorithm

## Declaration of Interests

The authors have no competing interests which may have influenced the work shown in this manuscript

## Contributions

K.J. designed the experiments, and B.C.K. designed the algorithms. All authors discussed the result and wrote the paper.